\documentclass[aps,pra]{revtex4}
\usepackage{graphicx}

\begin{document}

\title{Generalizing Several Theoretical Deterministic Secure Direct Bidirectional Communications to
Improve Their Capacities\\
\thanks{*Email: zhangzj@wipm.ac.cn }}

\author{Z. J. Zhang and Z. X. Man \\
{\normalsize Wuhan Institute of Physics and Mathematics, Chinese
Academy of Sciences, Wuhan 430071, China } \\
{\normalsize *Email: zhangzj@wipm.ac.cn }}

\date{\today}
\maketitle

\begin{minipage}{380pt}
Several theoretical Deterministic Secure Direct Bidirectional
Communication protocols are generalized to improve their
capacities by introducing the superdense-coding in the case
of high-dimension quantum states.\\

PACS Number(s): 03.67.Hk, 03.65.Ud\\
\end{minipage}

Recently, we propose three theoretical Deterministic Secure Direct
Bidirectional Communication protocols [1-3]. All these protocols
can be generalized to improve their capacities by introducing the
superdense-coding in the case of high-dimension states [4-5].

This work is supported by the National Natural Science Foundation
of China under Grant No. 10304022. \\

\noindent[1] Zhanjun Zhang, quant-ph/0403186.

\noindent[2] Z. J. Zhang and Z. X. Man, quant-ph/0403215.

\noindent[3] Z. J. Zhang and Z. X. Man, quant-ph/0403217.

\noindent[4] X. S. Liu, G. L. Long, D. M. Tong and Feng Li, Phys.
Rev. A {\bf65}, 022304 (2002).

\noindent[5] A. Grudka and A. Wojcik, Phys. Rev. A {\bf66}, 014301
(2002).

\end{document}